\RequirePackage{ifpdf}
\ifpdf 
\documentclass[pdftex]{sigma}
\else
\documentclass{sigma}
\fi

\usepackage{amsbsy,eucal}

\newcommand{\im}{\mathop{\rm i}}

\newcommand{\B}[1]{\boldsymbol{#1}}
\newcommand{\s}[1]{\textbf{\sffamily{#1}}}

\begin{document}
\allowdisplaybreaks

\renewcommand{\PaperNumber}{029}

\FirstPageHeading

\ShortArticleName{Large-$j$ Expansion Method for Two-Body Dirac
Equation}

\ArticleName{Large-$\boldsymbol{j}$ Expansion Method \\ for
Two-Body Dirac Equation}

\Author{Askold DUVIRYAK}

\AuthorNameForHeading{A. Duviryak}

\Address{Institute~for~Condensed~Matter~Physics of National
Academy of Sciences of Ukraine, \\ 1 Svientsitskii~Str., Lviv,
79011 Ukraine}

\Email{\href{mailto:duviryak@ph.icmp.lviv.ua}{duviryak@ph.icmp.lviv.ua}}

\ArticleDates{Received December 01, 2005, in f\/inal form February
15, 2006; Published online February 28, 2006}

\Abstract{By using symmetry properties, the two-body Dirac
equation in coordinate representation is reduced to the coupled
pair of radial second-order dif\/ferential equations. Then the
large-$j$ expansion technique is used to solve a bound state
problem. Linear-plus-Coulomb potentials of dif\/ferent spin
structure are examined in order to describe the asymptotic
degeneracy and f\/ine splitting of light meson spectra. }


\Classification{81Q05; 81Q15; 81Q20}

\section{Introduction}

    Two-body Dirac equations (2BDE), i.e., the Breit
equations \cite{Bre29} and its generalizations
\cite{B-K85,B-U87,Gra91,D-L96,Dar98,D-D02,D-D05,F-N90,N-F91}, are
used frequently for the description of relativistic bound state
problem, especially in nuclear \cite{S-T01,S-T04} and hadronic
\cite{Kro76,Chi87,Bra87,Tsi00} physics. Apart from two
free-particle Dirac terms, the 2BDE may include potentials which
are local matrix-functions in the coordinate representation. This
form provides an intuitive understanding of the interaction and
may suggest a proper physical choice of the potential in
phenomenological models.

    But the 2BDE are pathological if certain interaction terms are not
treated perturbationally. The set of radially reduced equations
\cite{Kro76,D-L96,D-D02} may possess non-physical energy-dependent
poles at f\/inite distance $r$ between particles
\cite{Khe82,CWW96}. Correspondingly, an exact boundary-value
problem becomes incorrect mathematically or improper for assumed
physical treatment.

    Here we consider a possibility to avoid pathological
peculiarities of 2BDE using a pseudo-perturbative technique
similar to $1/N$ \cite{M-S84,IPS84,Vak02} or $1/\ell$ expansions
\cite{M-B97,M-B98}. These methods are applicable to the case of a
strong coupling and are little af\/fected by boundary
peculiarities of the boundary-value problem.

    In our case natural expansion parameter is $1/j$, where $j$ is the
conserved total angular momentum. After the radial reduction is
performed, the 2BDE takes the form of the set of eight coupled
f\/irst-order dif\/ferential equations \cite{D-L96,D-D02}. Using a
chain of transformations we reduce it to the pair of coupled
second-order equations and apply the $1/j$ expansion technique.
The method is applied to the potential model of meson based on the
2BDE.

\section{2-body Dirac equation and its radial reduction}

In the centre-of-mass reference frame the 2BDE has the form:
\begin{gather}\label{2.1}
\left\{h_1(\B p) + h_2(-\B p) + U(\B r) - E\right\}F(\B r) = 0,
\end{gather}
where $F(\B r)$ is a 16-component wave function,
\begin{gather*}
h_a(\B p) =\B\alpha_a\cdot\B p + m_a\beta_a
\equiv{}-\im\B\alpha_a\cdot\B\nabla + m_a\beta_a, \qquad a=1,2,
\end{gather*}
are Dirac Hamiltonians of free fermions of mass $m_a$ and $U(\B
r)$ is an interaction potential. If $F(\B r)$ is presented in
$4\times4$-matrix representation, the operators $\B\alpha_a$ and
$\beta_a$ act as follows: $\B\alpha_1F=\B\alpha F$,
$\B\alpha_2F=F\B\alpha^{\rm T}$ etc, where $\B\alpha$ and $\beta$
are Dirac matrices.

    The potential $U(\B r)$ is the Hermitian matrix-function, it is invariant
under rotation and space inversion transformations (so that the
total Hamiltonian $H=h_1+h_2+U$ is too). Its general form is
parametrized by 48 scalar function of $r=|\B r|$ \cite{N-F91}. Of
physical meaning are potentials admitting f\/ield-theoretical
interpretation of interaction. In particular, potentials
ref\/lecting a~spin structure of vector and scalar relativistic
interactions are used frequently in potential quark models of
mesons. We will consider such a model in the Section~5 using few
examples of scalar and vector potentials known in a literature. In
the present section the structure of potential is not essential.

    In order to apply a pseudoperturbative expansion method to the 2-body
Dirac equation let us transform it to an appropriate form.

    First of all we perform a radial reduction. Following
    \cite{D-L96,D-D02} we put the wave eigenfunction $F(\B r)$ of the total
angular momentum $j$ and the parity $P$ into the $2\times 2$
block-matrix form:
\begin{gather}\label{2.12}
F(\B r) = \frac1r\left[
\begin{array}{cc}
\im s_1(r)\phi^A(\B n) + \im s_2(r)\phi^0(\B n) &
t_1(r)\phi^-(\B n) + t_2(r)\phi^+(\B n) \vspace{1mm}\\
u_1(r)\phi^-(\B n) + u_2(r)\phi^+(\B n) & \im v_1(r)\phi^A(\B n) +
\im v_2(r)\phi^0(\B n)
\end{array}
\right]
\end{gather}
for the parity $P=(-)^{j\pm1}$ states, and into a similar form for
the parity $P=(-)^{j}$ states but with superscripts interchanged
as follows: $(A,0)\leftrightarrow(-,+)$. Here $\B n=\B r/r$, the
bispinor harmo\-nics~$\phi^A(\B n)$ corresponds to a singlet state
with a total spin $s=0$ and an orbital momentum $\ell=j$,
and~$\phi^0(\B n)$, $\phi^-(\B n)$, $\phi^+(\B n)$ correspond to
triplet with $s=1$ and $\ell=j,j+1,j-1$. Then for $j>0$ the
eigenstate problem \eqref{2.1} reduces to the set of eight
f\/irst-order dif\/ferential equations with the functions
$s_1(r),\ldots, v_2(r)$ and the energy $E$ to be found.

    It is convenient to present this set in the following matrix form.
Let us introduce the 8-dimensional vector-function: $\s X(r) =
\{s_1(r), s_2(r), t_1(r),\dots, v_2(r)\}$. Then the set of radial
equations reads:
%
\begin{gather}\label{2.15}
\left\{\s H(j) \frac{d}{dr} + \s V(r,E,j)\right\}\s X(r) = E\s
X(r),
\end{gather}
where the $8\times 8$ real matrices $\s H(j)$ and $\s V(r,E,j)=\s
G(j)/r+\s m+\s U(r,j) - E\s I$ possesses properties $\s H^{\rm
T}=-\s H$, $\s V^{\rm T}=\s V$, the diagonal matrix $\s m={\rm
diag}(m_+\mathcal I,m_-\mathcal I,-m_-\mathcal I,-m_+\mathcal I)$
(here $\mathcal I$ is 2$\times$2 unit matrix and $m_\pm=m_1\pm
m_2$) and $j$- and $P$-dependent matrices $\s H(j)$, $\s G(j)$ are
constant (i.e., free of $r$), and matrix-potential $\s U(r,j)$
comes from interacting term of the equation \eqref{2.1}. For the
case $j=0$ components $s_2=t_2=u_2=v_2=0$ so that the dimension of
the problem \eqref{2.15} reduces from~8 to~4.

    It turns out that ${\rm rank}\,\s H=4$ (2 for $j=0$). In other words,
only four equations of the set~\eqref{2.15} are dif\/ferential
while remaining ones are algebraic. They can be split by means of
some orthogonal (i.e., of O(8) group) transformation. In new basis
we have
\begin{gather*}
\s X(r) = \left[
\begin{array}{c}
\s X_1 \\ \s X_2
\end{array}
\right], \qquad \s H = 2\left[
\begin{array}{cc}
\s J^{(2)} & \s 0 \\
\s 0 & \s 0
\end{array}
\right],
\end{gather*}
where $\s J^{(2)}$ is the symplectic $4 \times 4$ matrix. Thus we
arrive at the set
\begin{gather}
2\s J^{(2)}\s X'_1+\s V_{11}\s X_1+\s V_{12}\s X_2=0, \label{2.17} \\
                   \s V_{21}\s X_1+\s V_{22}\s X_2=0. \label{2.18}
\end{gather}
Eliminating $\s X_2$ from \eqref{2.17} by means of \eqref{2.18} we
get a dif\/ferential set for the 4-vector $\s X_1$
\begin{gather*}
\left\{\s J^{(2)} \frac{d}{dr}+\s V^\bot(r,E,j)\right\}\s
X_1(r)=0, \qquad{\rm where}\quad \s V^\bot=(\s V_{11}-\s V_{12}\s
V_{22}^{-1}\s V_{21})/2
\end{gather*}
while $\s X_2$ then follows from the algebraic relation $\s
X_2=-\s V_{22}^{-1}\s V_{21}\s X_1$.

    The elimination of $\s X_2$ causes
non-physical energy-dependent singular points (apart of $r=0$ and
physical singularities of potentials) in matrix elements of $\s
V^\bot$.

    Now we present the 4-vector $\s X_1$ in $2+2$ block form,
\begin{gather*}
\s X_1(r) = \left[
\begin{array}{c}
\Phi_1 \\ \Phi_2
\end{array}
\right], \qquad \s V^\bot = \left[
\begin{array}{cc}
\mathcal V_{11} & \mathcal V_{12} \\
\mathcal V_{21} & \mathcal V_{22}
\end{array}
\right],
\end{gather*}
eliminate then $\Phi_2$ and arrive at the second-order
dif\/ferential equations for 2-vector $\Phi_1$:
\begin{gather}\label{2.21}
\mathcal L(E)\Phi_1= \left\{\left(\frac{d}{dr}+\mathcal
V_{12}\right)\left[\mathcal V_{22}\right]^{-1}
\left(\frac{d}{dr}-\mathcal V_{21}\right)+\mathcal
V_{11}\right\}\Phi_1=0.
\end{gather}

    The matrix $\mathcal V_{22}$ is diagonal for all potentials considered
in Section 5 (and many other ones). In these cases we can perform
the transformation:
\begin{gather*}
\tilde\Phi_1=\Phi_1/\sqrt{\mathcal V_{22}}, \qquad \tilde{\mathcal
L}=\sqrt{\mathcal V_{22}}{\mathcal L}\sqrt{\mathcal V_{22}}
\end{gather*}
providing for the operator $\tilde{\mathcal L}$ the form which is
as close as possible to 2-term form:
\begin{gather}\label{2.23}
\tilde{\mathcal L}(E)=\frac{d^2}{dr^2}- \mathcal W(r,E,j)
-\left\{Z(r,E,j),\frac{d}{dr}\right\}_+ \mathcal J^{(1)};
\end{gather}
here $\mathcal W(r,E,j)$ is a symmetric $2\times 2$ matrix,
$\mathcal J^{(1)}$ is $2\times 2$ symplectic matrix and
$\{\cdot,\cdot\}_+$ denotes the anticommutator.

 We are going to apply the $1/j$ expansion method to the equation
\eqref{2.23}. In many physically interesting cases the function
$Z$ vanishes or it is negligible at $j$ large. Thus the wave
equation has a~$2\times 2$ matrix 2-term form which is convenient
for application of the method. In other cases the third term of
the operator \eqref{2.23} contains a f\/irst-order derivative via
of\/f-diagonal matrix elements only. This form is tractable too,
but with more tedious calculations. We do not consider such
equations in this paper. Before proceeding further, we study a
simpler example of a single 2-term relativistic equation.

\section[Todorov equation via $1/l$ method]{Todorov equation via $\boldsymbol{1/\ell}$ method}

    Here we consider the Todorov-type equation describing the
relativistic system of two interacting scalar particles in the
centre-of-mass reference  frame \cite{Tod71,RST85,Duv02}:
\begin{gather*}
\left\{\B p^2 + U(r,E) - b(E)\right\}\Psi(\B r) = 0.
\end{gather*}
Here $\B p = -\im\B\nabla$, the quasipotential $U(r,E)$ depends on
energy $E$ of the system, and the binding parameter $b(E)$ is the
following function of $E$,
\begin{gather}\label{3.2}
b(E) = \tfrac 14 E^2 - \tfrac 12 \big(m_1^2+m_2^2\big) + \tfrac 14
\big(m_1^2-m_2^2\big)^2/E^2,\quad\ {\rm so~that}\quad\ E(b) =
\sum_{a=1}^2\sqrt{m_a^2+b}.
\end{gather}
The corresponding radial equation takes the form
\begin{gather}\label{3.3}
\left\{\frac{d^2}{dr^2} - W(r,E,\ell)\right\}\Psi(r) = 0,
\end{gather}
where $\ell$ is the angular momentum quantum number, and
\begin{gather}\label{3.4}
W(r,E,\ell)=U(r,E)+\ell(\ell+1)/r^2-b(E).
\end{gather}

    Let us consider motion of the system in the
neighbourhood of classical stable circular orbit. Given $\ell>0$,
the radius $r_{\rm c}=r_{\rm c}(\ell)$ of the stable circular
orbit and the corresponding energy $E_{\rm c}=E_{\rm c}(\ell)$
satisfy conditions:
\begin{gather}\label{3.5}
W(r_{\rm c},E_{\rm c},\ell)=0,\qquad
\partial W(r_{\rm c},E_{\rm c},\ell)/\partial r_{\rm c}=0
\end{gather}
and $\partial^2 W(r_{\rm c},E_{\rm c},\ell)/\partial r_{\rm
c}^2>0$; here $\partial W(r_{\rm c},E_{\rm c},\ell)/\partial
r_{\rm c} \equiv
\partial W(r,E,\ell)/\partial
r\,\smash{\rule[-0.55em]{0.5pt}{1.4em}}_{r=r_{\rm c}\atop E=E_{\rm
c}} \rule[-0.8em]{0.0pt}{1.4em}$ etc.

One puts $r=r_{\rm c}+\Delta r$ and $E=E_{\rm c}+\Delta E$ where
$\Delta r$ and $\Delta E$ are small in some meaning, and expand
the function $W(r_{\rm c}+\Delta r, E_{\rm c}+\Delta E, \ell)$ in
power series with respect $\Delta r$ and $\Delta E$. Then due to
the conditions \eqref{3.5} the leading terms of this expansion
represent the harmonic oscillator potential and other ones are
anharmonic terms. If the conditions \eqref{3.5} hold for any large
value of $\ell$ it is possible by renormalization of $\Delta r$
and $\Delta E$ to single out in the equation \eqref{3.3} the
$\ell$-independent harmonic oscillator problem and anharmonic
perturbations which disappear if $\ell\to\infty$. This is the idea
of $1/\ell$ expansion method. Application of pseudoperturbative
techniques of this type \cite{M-S84,IPS84,Vak02,M-B97,M-B98} to
our case meets two peculiarities: the equation \eqref{3.3}
represents a nonlinear spectral problem, and an exact solution of
the equations \eqref{3.5} may appear to be unknown or too
cumbersome for practical use. Thus we modify the technique.

    Let us introduce the parameter $\lambda=1/\sqrt\ell$ which is small
at $\ell$ large. Since the exact form of the functions $r_{\rm
c}(\ell)$ and $E_{\rm c}(\ell)$ is unknown in general, we f\/irst
determine asymptotics $r_{\rm c}\sim r_\infty(\lambda)$, $b_{\rm
c}=b(E_{\rm c}) \sim b_\infty(\lambda)$ at $\lambda\to0$ which may
be found much easier. Then the functions  $r_{\rm c}(\ell)$
and~$E_{\rm c}(\ell)$ can be presented in the form:
\begin{alignat}{3}
&r_{\rm c}(\lambda)= r_\infty(\lambda)\rho(\lambda),&& b_{\rm
c}(\lambda)=b_\infty(\lambda)\mu(\lambda), &
\nonumber\\
&\rho(\lambda)=1+\lambda\rho^{(1)}+\lambda^2\rho^{(2)}+\cdots,\qquad
&&
\mu(\lambda)=1+\lambda\mu^{(1)}+\lambda^2\mu^{(2)}+\cdots,\label{3.8}
\end{alignat}
where expansion coef\/f\/icients $\rho^{(n)}$, $\mu^{(n)}$,
$n=1,2,\dots$ (and thus the analytical functions $\rho(\lambda)$
and $\mu(\lambda)$) can be found, step by step, from the
conditions:
\begin{gather}\label{3.12}
\bar W(\rho,\mu,\lambda)=0,\qquad
\partial \bar W(\rho,\mu,\lambda)/\partial\rho=0
\end{gather}
and $\partial^2 \bar W(\rho,\mu,\lambda)/\partial\rho^2>0$; here
the dimensionless function $\bar W(\rho,\mu,\lambda)$ is
constructed by the direct use of \eqref{3.8} in \eqref{3.4} and
normalizing in order that $\bar W(\rho,\mu,\lambda)$ to be regular
at $\lambda\to 0$,
\begin{gather*}
\bar W(\rho,\mu,\lambda)=
\lambda^4r_\infty^2W\left[r_\infty\rho,E(b_\infty\mu),1/\lambda^2\right].
\end{gather*}

    Now we go to the dimensionless variable $r\to\xi$ and spectral
parameter $b(E)\to\epsilon$,
\begin{gather}\label{3.10}
r=r_\infty(\lambda)[\rho(\lambda)+\lambda\xi], \qquad
b=b_\infty(\lambda)\left[\mu(\lambda)+\lambda^2\epsilon\right],
\end{gather}
in terms of which the equation \eqref{3.3} takes the form
\begin{gather}\label{3.15}
\left\{\frac{d^2}{d\xi^2} -
\frac1{\lambda^2}w(\xi,\epsilon,\lambda)\right\}\psi(\xi) = 0
\end{gather}
with
\begin{gather*}
\psi(\xi)=\Psi[r_{\infty}(\rho+\lambda\xi)],
\end{gather*}
and
\begin{gather}\label{3.15a}
w(\xi,\epsilon,\lambda)= \bar
W(\rho+\lambda\xi,\,\mu+\lambda^2\epsilon,\,\lambda).
\end{gather}
If the functions $\rho(\lambda)$ and $\mu(\lambda)$ satisfy the
conditions \eqref{3.12}, the equation \eqref{3.15} is nonsingular
at $\lambda\to0$. This is true even if we use the f\/irst-order
approximate solution to \eqref{3.12} in \eqref{3.10},
\begin{gather}\label{3.19a}
\rho(\lambda)=1+\lambda\rho^{(1)}, \qquad
\mu(\lambda)=1+\lambda\mu^{(1)}.
\end{gather}
Indeed, using the notation $\partial \bar W^{(0)}/\partial\mu=
\lim\limits_{\lambda\to0}\partial \bar W/\partial\mu=
\partial \bar W/\partial\mu(\rho=1,\mu=1,\lambda=0)$ etc. we have
\begin{gather}
\frac1{\lambda^2}w(\xi,\epsilon,\lambda)= \frac1{\lambda^2}\bar
W\left[  \rho(\lambda)+\lambda  \xi,\,
  \mu(\lambda)+\lambda^2  \epsilon,\,\lambda\right]
 \nonumber\\
\phantom{\frac1{\lambda^2}w(\xi,\epsilon,\lambda)}{}
=\frac{1}{\lambda^2}\bar
W^{(0)}+\frac{1}{\lambda}\left\{\frac{\partial \bar
W^{(0)}}{\partial\rho}\big(\rho^{(1)}+ \xi\big) + \frac{\partial
\bar W^{(0)}}{\partial\mu}\big(\mu^{(1)}+\lambda
\epsilon\big)+\frac{\partial \bar
W^{(0)}}{\partial\lambda}\right\}  \nonumber\\
\phantom{\frac1{\lambda^2}w(\xi,\epsilon,\lambda)=}{} +
\frac12\frac{\partial^2 \bar
W^{(0)}}{\partial\rho^2}\big(\rho^{(1)}+ \xi\big)^2 +
\frac12\frac{\partial^2 \bar
W^{(0)}}{\partial\mu^2}\big[\mu^{(1)}\big]^2 +
\frac12\frac{\partial^2 \bar
W^{(0)}}{\partial\lambda^2}  \nonumber\\
\phantom{\frac1{\lambda^2}w(\xi,\epsilon,\lambda)=}{} +
\frac{\partial^2 \bar W^{(0)}}{\partial\rho\partial\mu}
\big(\rho^{(1)}+ \xi\big)\mu^{(1)} + \frac{\partial^2 \bar
W^{(0)}}{\partial\rho\partial\lambda} \big(\rho^{(1)}+ \xi\big) +
\frac{\partial^2 \bar W^{(0)}}{\partial\mu\partial\lambda}
\mu^{(1)} + O(\lambda). \label{3.22}
\end{gather}
Singular terms are absent if the following set of equations holds:
\begin{gather}
\bar W^{(0)}=0, \qquad \partial
\bar W^{(0)}/\partial\rho=0, \label{3.23}\\
\frac{\partial \bar W^{(0)}}{\partial\mu}\mu^{(1)}+\frac{\partial
\bar W^{(0)}}{\partial\lambda}=0. \label{3.24}
\end{gather}
Besides, zero-order terms which are linear in $ \xi$ disappear if
\begin{gather}\label{3.25}
\frac{\partial^2 \bar W^{(0)}}{\partial\rho^2}\rho^{(1)} +
\frac{\partial^2 \bar W^{(0)}}{\partial\rho\partial\mu}\mu^{(1)} +
\frac{\partial^2 \bar W^{(0)}}{\partial\rho\partial\lambda}=0.
\end{gather}
Notice that the equations \eqref{3.23} and
\eqref{3.24}--\eqref{3.25} represent the conditions \eqref{3.12}
in the zeroth and f\/irst orders of $\lambda$, respectively. Thus
the equations \eqref{3.23} hold identically and
\eqref{3.24}--\eqref{3.25} are linear equations with $\rho^{(1)}$
and $\mu^{(1)}$ to be found.

    In zero-order approximation the equation \eqref{3.15} reduces to the
harmonic oscillator problem
\begin{gather}\label{3.26}
\left\{\frac{d^2}{d\xi^2} + \kappa \epsilon - \nu -
\omega^2\xi^2\right\}\psi(\xi) = 0
\end{gather}
with
\begin{gather}
\kappa=-\frac{\partial \bar W^{(0)}}{\partial\mu}, \qquad
\omega^2=\frac12\frac{\partial^2 \bar W^{(0)}}{\partial\rho^2},\label{3.18}\\
\nu= -\frac12\frac{\partial^2 \bar
W^{(0)}}{\partial\rho^2}\big[\rho^{(1)}\big]^2 +
\frac12\frac{\partial^2 \bar
W^{(0)}}{\partial\mu^2}\big[\mu^{(1)}\big]^2 +
\frac12\frac{\partial^2 \bar W^{(0)}}{\partial\lambda^2} +
\frac{\partial^2 \bar W^{(0)}}{\partial\mu\partial\lambda}
\mu^{(1)}, \label{3.27}\\
\mu^{(1)}= -\frac{\partial^2 \bar W^{(0)}/\partial\lambda}
{\partial^2 \bar W^{(0)}/\partial\mu}, \qquad \rho^{(1)} =
-\frac1{\partial^2 \bar W^{(0)}/\partial\rho^2}
\left\{\frac{\partial^2 \bar
W^{(0)}}{\partial\rho\partial\mu}\mu^{(1)} + \frac{\partial^2 \bar
W^{(0)}}{\partial\rho\partial\lambda}\right\}. \label{3.28}
\end{gather}
The higher-order terms in the expansion \eqref{3.22} can be
considered as perturbations of the oscillator problem
\eqref{3.26}. They depend, in general, on the spectral parameter
$\epsilon$ and can be taken into account by means of the
perturbative procedure \cite{RST85} is appropriate to this case.
Otherwise the treatment is similar to
\cite{M-S84,IPS84,Vak02,M-B97,M-B98}.

    The eigenvalues in zero-order approximation
$ \epsilon_{n_r}=[\omega(2n_r+1)+\nu]/\kappa$, where
$n_r=0,1,\dots$ is a radial quantum number, are to be corrected by
means of higher orders of perturbative procedure. Then, using of
the 2nd equation of \eqref{3.10} in \eqref{3.2} gives us the
energy spectrum.

\section[Breit-type equation via $1/j$ method]{Breit-type equation via $\boldsymbol{1/j}$ method}

    At this point we return to the radial 2BDE in the form
$\tilde{\mathcal L}(E)\tilde\Phi_1=0$, where the $2\times 2$
matrix operator $\tilde{\mathcal L}(E)$ is given by equation
\eqref{2.23} with the last term neglected. Let us put
\begin{gather*}
\Phi_1 = \left[
\begin{array}{c}
\Psi_1 \\ \Psi_2
\end{array}
\right],
\end{gather*}
where $\Psi_1$ and $\Psi_2$ are components of $\Phi_1$. Then the
equation \eqref{2.21} can be presented as a pair of coupled
ordinary second-order dif\/ferential equations:
\begin{gather}
\frac{d^2}{dr^2}\Psi_1(r) - W_1(r,E,j)\Psi_1(r) =
Y(r,E,j)\Psi_2(r),
\label{4.2}\\
\frac{d^2}{dr^2}\Psi_2(r) - W_2(r,E,j)\Psi_2(r) =
Y(r,E,j)\Psi_1(r). \label{4.3}
\end{gather}
We will treat this system perturbationally using the pseudosmall parameter $\lambda=1/\sqrt{j}$.

    Let us suppose for a moment that the right-hand side of the system
\eqref{4.2}--\eqref{4.3} can be ignored, so that these equations
decouple. Then we can apply to each of the equations the scheme of
the Section~3.  We def\/ine radii and energies of circular orbits
by means of the conditions:
\begin{gather*}
W_i(r_i,E_i,j)=0,\qquad \frac{\partial W_i(r_i,E_i,j)}{\partial
r}=0,\qquad \frac{\partial^2 W_i(r_i,E_i,j)}{\partial
r^2}>0,\qquad i=1,2.
\end{gather*}
Then we single out asymptotics of these functions of $\lambda$ by
means of the relations:
\begin{alignat*}{3}
& r_i(\lambda)= r_{i\infty}(\lambda)\rho_i(\lambda),&&
b_i(\lambda)=b_{i\infty}(\lambda)\mu_i(\lambda),&\\
&
\rho_i(\lambda)=1+\lambda\rho_i^{(1)}+\lambda^2\rho_i^{(2)}+\cdots,\qquad&&
\mu(\lambda)=1+\lambda\mu_i^{(1)}+\lambda^2\mu_i^{(2)}+\cdots,&
\end{alignat*}
and, using the relations
\begin{gather}\label{4.7}
r=r_{i\infty}(\lambda)[\rho_i(\lambda)+\lambda\xi_i], \qquad
b=b_{i\infty}(\lambda)[\mu_i(\lambda)+\lambda^2\epsilon_i],\qquad
i=1,2
\end{gather}
we reformulate the equation \eqref{4.2} in terms of the
dimensionless variable $\xi_1$ and the spectral parameters
$\epsilon_1$ while the equation \eqref{4.3} -- in terms of $\xi_2$
and $\epsilon_2$. Finally, we perform expansion of the equations
into powers of $\lambda$ and solve them separately.

    Now we are going to take actual coupling of the equations \eqref{4.2}
and \eqref{4.3} into account. First of all, we note that the
variables $\xi_1$ and $\xi_2$ are not  of one another, and the
spectral parameters~$\epsilon_1$ and $\epsilon_2$ are also not
independent. Thus we should choose common variables in both
equations.

    Let us f\/irst choose $\xi=\xi_1$, $\epsilon=\epsilon_1$. Then the set
\eqref{4.2}--\eqref{4.3} reduces to the form:
\begin{gather}
\psi''_1(\xi) -\frac1{\lambda^2}w_1(\xi,
\epsilon,\lambda)\psi_1(\xi)=
y(\xi,\epsilon,\lambda)\psi_2(\xi),\label{4.8}\\
\psi''_2(\xi) -\frac1{\lambda^2}w_2(\xi,
\epsilon,\lambda)\psi_2(\xi)=
y(\xi,\epsilon,\lambda)\psi_1(\xi),\label{4.9}
\end{gather}
where
\begin{gather}
\psi_i(\xi)=\Psi_i[r_{1\infty}(\rho_1+\lambda\xi)], \qquad
i=1,2,\label{4.10}\\
w_i(\xi,\epsilon,\lambda)=
\lambda^4r_{1\infty}^2W_i\left[r_{1\infty}(\rho_1+\lambda\xi),
E\!\left(b_{1\infty}(\mu_1+\lambda^2\epsilon)\right),1/\lambda^2\right],
\label{4.11}\\
y(\xi,\epsilon,\lambda)=
\lambda^2r_{1\infty}^2Y\left[r_{1\infty}(\rho_1+\lambda\xi),
E\!\left(b_{1\infty}(\mu_1+\lambda^2\epsilon)\right),1/\lambda^2\right].
\label{4.12}
\end{gather}

    The functions \eqref{4.11}--\eqref{4.12} are regular
at $\lambda\to0$. Moreover, the general structure of the function
$w_1$ is the same as that of $w$ in the Section 3 (see
equations~\eqref{3.15a}, \eqref{3.22}). In particular,
$w_1=O(\lambda^2)$. Thus the equation \eqref{4.8} is similar
to~\eqref{3.15} (but with non-zero right-hand side). It admits
similar expansion in $\lambda$.

    On the contrary, the function $w_2$ may have a dif\/ferent behaviour
at $\lambda\to0$. Here we consider three cases.

    {\bf 1.} Let $r_{2\infty}\ne r_{1\infty}$ and
$b_{2\infty}\ne b_{1\infty}$. Then $w_2=O(\lambda^{-n})$, $n\ge0$
(except perhaps very special examples which we do not consider).
In this case one can solve formally the equation \eqref{4.9} in
favour of $\psi_2(\xi)$ as follows:
\begin{gather}\label{4.13}
\psi_2=
-\left(1-\frac{\lambda^2}{w_2}\frac{\partial^2}{\partial\xi^2}\right)^{-1}
\frac{\lambda^2}{w_2}y\psi_1 = -\sum\limits_{n=0}^{\infty}
\left(\frac{\lambda^2}{w_2}\frac{\partial^2}{\partial\xi^2}\right)^n
\frac{\lambda^2}{w_2}y\psi_1.
\end{gather}
This representation leads to the loss of solutions for $\psi_2$
which are not analytical in $\lambda$ and thus have nothing to do
with the perturbation procedure. The use of \eqref{4.13} in the
r.h.s. of \eqref{4.8} permits us to eliminate $\psi_2$ from
\eqref{4.8} and thus to obtain a close wave equation for $\psi_1$.
The structure and treatment of this equation are the same as those
of the equation \eqref{3.15}. Moreover, it is obvious from
\eqref{4.13} that at least the zero- and the f\/irst-order terms
of $\psi_2$ vanish.  Thus the r.h.s.\ of \eqref{4.8} does not
contribute in lower orders of perturbation procedure. In
zero-order approximation we have the oscillator problem.

    {\bf 2.} Let $r_{2\infty}=r_{1\infty}$ and $b_{2\infty}=b_{1\infty}$
but $\rho_2-\rho_1=O(\lambda)$ and $\mu_2-\mu_1=O(\lambda)$. Then
$w_2=O(\lambda)$. Since $\lambda^2/w_2=O(\lambda)$ the
perturbative treatment \eqref{4.13} of the equation \eqref{4.9} is
still valid. The only dif\/ference from the case {\bf 1} is that
the r.h.s. of equation \eqref{4.8} may contribute in the f\/irst
order of~$\lambda$.

    In both the above cases we used the dimensionless variable $\xi_1$
and obtained a closed eigenstate equation (which we will reference
to as the problem 1) for the wave function $\psi_1(\xi_1)$ and the
spectral parameter $\epsilon_1$. We can proceed with the variable
$\xi_2$ and obtain the problem~2 for the function $\psi_2(\xi_2)$
and the parameter $\epsilon_2$. One might be inclined to think
that both problems~1 and~2 are equivalent and lead to the same
spectrum (in terms of energy $E$). Actually, dif\/ferent problems
complement one another. This is evident from equation \eqref{4.7}
leading to the relation:
\begin{gather*}
\epsilon_2-\epsilon_1=\frac1{\lambda^2}\left\{\frac{b_{1\infty}}{b_{2\infty}}
\mu_1-\mu_2\right\} + \left\{\frac{b_{1\infty}}{b_{2\infty}}
-1\right\}\epsilon_1.
\end{gather*}
Indeed, in both {\bf 1} and {\bf 2} cases
$|\epsilon_2-\epsilon_1|\to\infty$ if $\lambda\to0$. It does mean
that an arbitrary energy level $E$ calculated by means of
eigenvalue $\epsilon^{(0)}_1$ of zero-order oscillator problem~1
(with the use of equations \eqref{4.7} and \eqref{3.2}) cannot be
obtained by means of any f\/inite eigenvalue $\epsilon^{(0)}_2$ of
the problem~2 and vice versa. Higher-order corrections to
$\epsilon^{(0)}_1$ (or $\epsilon^{(0)}_2$) are small and do not
change qualitatively this picture. Thus dif\/ferent problems
generate dif\/ferent branches of the energy spectrum of the
original set of equation. In this respect the following special
case dif\/fers essentially from the previous ones.

    {\bf 3.} Let $r_{2\infty}=r_{1\infty}$ and $b_{2\infty}=b_{1\infty}$
but $\rho_2-\rho_1=O(\lambda^n)$ and $\mu_2-\mu_1=O(\lambda^n)$,
$n\ge2$. Then $w_2=O(\lambda^2)$. Both equations \eqref{4.8} and
\eqref{4.9} have similar structure and should be treated on the
same footing. Use of common variables $\xi$, $\epsilon$ def\/ined
by \eqref{3.10} and \eqref{3.19a} is appropriate to this case. In
the zero-order approximation we obtain the coupled pair of wave
equations (on the contrary to the cases {\bf 1} and {\bf 2}
where we had a single wave equation). In physically meaningful
cases (of Section~5, for example) they have the form:
\begin{gather}
\left\{d^2/d\xi^2 + \kappa\epsilon - \nu_1 -
\omega^2\xi^2\right\}\psi_1(\xi) = \chi\psi_2(\xi),
\label{4.15}\\
\left\{d^2/d\xi^2 + \kappa\epsilon - \nu_2 -
\omega^2\xi^2\right\}\psi_2(\xi) = \chi\psi_1(\xi), \label{4.16}
\end{gather}
where $\chi=\lim\limits_{\lambda\to0}y={\rm const}$, and
parameters $\nu_i$, $\kappa$ and $\omega$ are related to the
functions $w_i$ ($i=1,2$) by the equation of the type of
\eqref{3.18}, \eqref{3.27} and \eqref{3.28}. The equations
\eqref{4.15}, \eqref{4.16} can be evidently reduced to the pair of
similar equations but with parameters
$\tilde\nu_i=\{\nu_1+\nu_2\pm\sqrt{(\nu_1-\nu_2)^2+4\chi^2}\}/2$
($i=1,2$) and $\tilde\chi=0$. Thus they become split equations of
the form \eqref{3.26}. The eigenvalues $\epsilon$ corresponding to
the f\/irst and second equations are separated by f\/inite
constant $\tilde\nu_1-\tilde\nu_2$. Thus the corresponding states
mix in higher orders of perturbation procedure.

\section{Application: Regge trajectories of mesons}

    Here we apply the pseudoperturbative treatment of 2BDE in meson
spectroscopy.

    It is known \cite{LSG91} that spectra of heavy mesons are
described well by the nonrelativistic potential model with
QCD-motivated funnel potential $u(r)=u_l(r)+u_C(r)$, where
\begin{alignat}{3}
& u_C(r)=-\alpha/r,\qquad && \alpha=0.27, \label{5.1}& \\
& u_l(r)= ar, &&  a=0.25\div0.3{\rm GeV}^2. \label{5.2}&
\end{alignat}
The Coulomb part \eqref{5.1} of this potential describes a
nonrelativistic limit of the vector one-gluon exchange interaction
while the linear part \eqref{5.2} is suggested by the area law in
the lattice approximation of QCD and has presumably scalar or
scalar-vector nature.

    Description of light meson spectroscopy needs application of
appropriate relativistic models. Most of them are related to the
string theory. From the theoretical viewpoint the most interesting
are QCD-motivated relativistic models embracing properties of both
 heavy and light mesons. Such models should ref\/lect the
scalar-vector structure of interaction and should lead to
funnel-type potential in the nonrelativistic limit.

    A natural candidate for the relativistic potential model is the2BDE with a short-range vector potential and a long-range
scalar one. At least three general structures of vector potential
are used in the literature,
\begin{gather}
U_{\rm v}(r)=u_{\rm v}(r), \label{2.3}\\
U_{\rm v}(r)=\{1-\B\alpha_1\cdot\B\alpha_2\}u_{\rm v}(r),
\label{2.4}\\
U_{\rm v}(\B r)=\big \{1-\tfrac 12
\B\alpha_1\cdot\B\alpha_2\big\}u_{\rm v}(r) + \tfrac 12 (\B
n\cdot\B\alpha_1)(\B n\cdot\B\alpha_2)ru'_{\rm v}(r), \label{2.5}
\end{gather}
with $u_{\rm v}(r)=u_C(r)$ or another short-range potential; here
$u'(r)=du(r)/dr$. The potential~\eqref{2.3} is only a static part
of vector interaction (see \cite{D-L96}). The relativistic vector
f\/ield kinematics is taken into account in the potential
\eqref{2.4} (see \cite{Tsi00,D-D05}) which, for the Coulomb case,
was f\/irst proposed by Eddington and Gaunt \cite{Edd29,Gau29}. In
the generalization \eqref{2.5} of the Breit potential \cite{Bre29}
retardation terms have been added \cite{D-D05}. Two dif\/ferent
scalar potentials,
\begin{gather}
U_{\rm s}(r)=\beta_1\beta_2u_{\rm s}(r), \label{2.6}\\
U_{\rm s}(r)=\tfrac 12 (\beta_1+\beta_2)u_{\rm s}(r), \label{2.7}
\end{gather}
come from dif\/ferent couplings of scalar mediating f\/ield with
fermionic f\/ields. The f\/irst one~\eqref{2.6} arises from the
Yukawa interaction (see \cite{Dar98}) while the second one
\eqref{2.7} corresponds to so called ``minimal'' coupling
\cite{Bra87}. The latter and also two following potentials can be
treated as static approximation of various QFT-motivated scalar
quasipotentials \cite{Sal52,Fau66,Khe69,Khe82}:
\begin{gather}
U_{\rm s}(r) =\tfrac 12 (1+\beta_1\beta_2)u_{\rm s}(r), \label{2.10}\\
U_{\rm s}(r)= \tfrac 14 (1+\beta_1)(1+\beta_2)u_{\rm s}(r).
\label{2.11}
\end{gather}

    The perturbative treatment of Breit-type
equations has been used for calculating a f\/ine splitting in
spectra of heavy mesons \cite{Chi87,Tsi00}. Light meson spectra
are essentially relativistic and need a nonperturbative statement
of the problem which is inconsistent because of non-physical
singularities of radial equations. To avoid these dif\/f\/iculties
in numerical calculations one is forced to invent sophisticated
potentials and impose rather artif\/icial boundary conditions
\cite{Bra87}.

    Using the pseudoperturbative treatment of 2BDE with dif\/ferent
combinations of potentials \eqref{5.1}--\eqref{2.11} we obtain
analytical expressions for meson mass spectra and estimate a role
of general structure and input parameters of potentials in the
model. We consider mass spectra of lightest mesons (containing $u$
and $d$ quarks only) and try to reproduce their following general
features:
\begin{enumerate}
\vspace{-1mm}\itemsep=0pt \item[i)] Mass spectra of light mesons
fall into the family of straight lines in the ($E^2,j$)-plane
known as Regge trajectories.

\item[ii)] Regge trajectories are parallel; slope parameter
$\sigma$ is an universal quantity, $\sigma=1.15{\rm
GeV}^2=(4\div4.5)a$.

\item[iii)] Nonrelativistic classif\/ication of light mesons as
$\left(n^{2s+1}\!\ell_j\right)$--states of quark-antiquark system
is adequate; i.e., radial quantum number $n_r=n-\ell-1$ enumerates
leading ($n_r=0$) and daughter ($n_r=1,2,\dots$) Regge
trajectories, spin $s=0,1$ corresponds to mass singlets and
triples etc.

\item[iv)] Spectrum is $\ell$$s$-degenerated, i.e., masses are
distinguished by $\ell$ (not by $j$) and $n_r$.

\item[v)] States of dif\/ferent $\ell$ possess an accidental
degeneracy which fact causes a tower structure of the spectrum.

\item[vi)] Hyperf\/ine $ss$-splitting is relatively small, about
$5\div6\,\%$ of $\sigma$.\vspace{-1mm}
\end{enumerate}

    For this purpose we use the nonrelativistic potential function
\eqref{5.1} and \eqref{5.2} in vector and scalar potentials of
dif\/ferent spin structure \eqref{2.3}--\eqref{2.11} and calculate
pseudoperturbative spectrum in zero-order approximation.
Classif\/ication of states then is done using singlet-triplet
properties of large-large component of wave function \eqref{2.12}
in the nonrelativistic limit.

    If the vector short-range interaction is ignored and scalar
potentials \eqref{2.6}--\eqref{2.11} are used with $u_{\rm
s}(r)=ar$ the pseudoperturbative mass (i.e., energy) of meson in
zero-order approximation has the following form:
\begin{gather}\label{5.3}
E_A^2=ka\big[\ell+\tfrac 12  + \eta\big(n_r+\tfrac 12 \big)\big] +
\zeta m_+\sqrt{2a\ell} + \delta_1m_+^2 - \delta_2m_1m_2 +
O\big(1/\sqrt{\ell}\big),\\
E_0^2 = E_A^2,\qquad E_\pm^2 = E_A^2 \pm \varkappa a; \nonumber
\end{gather}
here $m_+=m_1+m_2$, and $k$, $\eta$, $\zeta$, $\delta_1$,
$\delta_2$, $\varkappa$ are dimensionless constants depending on
the potential chosen.

    Four families of energy levels $E_i$ ($i=A,0,-,+$) form
trajectories in the ($E^2,\ell$)-plane which are nearly straight.
Indeed, $\zeta=0\div2$ for all the potentials considered and rest
masses $m_a$ \mbox{$(a=1,2)$} of lightest mesons are small
compared to the energy scale $\sqrt{\sigma}$. Thus the parameter
$\zeta m_+\sqrt{a}$ determining a curvature of trajectories is
small. Parameters $\delta_1$ and $\delta_2$ determine a~common
shift of all the trajectories and are not important for the
present discussion. Below  we discuss the calculated values of
parameters $k$, $\varkappa$ and $\eta$ determining the slope of
trajectories and their degeneracy properties.

    In the \eqref{2.6} case $k=4$ so that the slope $\sigma=ka$ matches quite
well to that of property~ii); $\eta=2$ causes  accidental
degeneracy typical for the harmonic oscillator; but $\varkappa=4$
leads to $j$-dependence of energy (not $\ell$-dependence) so that
the $ls$-degeneracy is absent.

    In the \eqref{2.7} case $k=4$ and $\eta=2$, so that the slope and
the accidental degeneracy are the same as in the \eqref{2.6} case;
$\varkappa=4-3\sqrt{2}\approx-0.243$ provides an approximate
$ls$-degeneracy, with accuracy $6\,\%$; the splitting is of order
of the actual $ss$-splitting  (see property vi)).

    In the \eqref{2.10} case $k=\varkappa=3\sqrt{3}\approx5.196$,
$\eta=\sqrt{3}\approx1.732$; none of these values match well to
properties ii)--vi).

    In the \eqref{2.11} case $k=\sqrt{23-\sqrt{17}}(7+\sqrt{17})^{2}/128
\approx4.2$ provides the best f\/it of $\sigma$ to that of
property ii);
$\eta=(\sqrt{17}-3)\sqrt{102+26\sqrt{17}}/8\approx2.03$ leads to
nearly precise oscillator-like degeneracy, with accuracy
$1.5\,\%$; $\varkappa=0$ provides exact $ls$-degeneracy.

    Taking into account the vector short-range interaction (one of
potentials \eqref{2.3}--\eqref{2.5} with $u_{\rm v}(r)=u_C(r)$)
results in a parallel shift of Regge trajectories. The value of
the shift is of the order $\alpha a$, it depends on the vector
potential chosen and is dif\/ferent (in general) for dif\/ferent
trajectories $E_i$ ($i=A,0,-,+$).

    It has been proved in the framework of single-particle Dirac equation
a possibility of conf\/i\-ne\-ment by means of vector and equally
mixed vector-scalar long-range interactions
\cite{L-R83,B-J82,HLS96}. We exa\-mi\-ned these cases in 2BDE
approach using dif\/ferent vector potentials
\eqref{2.3}--\eqref{2.5} with $u_{\rm v}(r)=ar$. Corresponding
zero-order pseudo-perturbative spectra have a form similar to
\eqref{5.3}. The dif\/ference is that $k=k_i=8\div12$ is  two times or
more larger than desired, and is dif\/ferent for $i=A,0,-,+$
(i.e., trajectories are not parallel).

\section{Summary}

    The Breit equation and its generalizations (2BDE) possess
non-physical singularities. In some cases these points lay far
from the physically important domain but they make a boundary
problem incorrect or physically improper \cite{Khe82,Tsi00}. In
order to avoid this dif\/f\/iculty and to use the 2BDE in the
relativistic bound state problem, especially for the case of
strong coupling, we develop the $1/j$ expansion method.

    The method is based on the large-$N$ or large-$\ell$ techniques
applicable to the radial Schr\"odinger equation. In our case the
2BDE is reduced to the coupled pair of quasipotential-type
equations which structure causes principal modif\/ication of known
techniques. Other changes  are related to the fact that the
equations represent a nonlinear spectral problem with cumbersome
quasipotentials.

    We apply this pseudoperturbative method to the 2BDE with the
linear+Coulomb potential of dif\/ferent scalar-vector structure.
In all cases in the zero-order approximation it was obtained the
Regge trajectories which are linear asymptotically. Linear
potentials of two scalar structures~\eqref{2.7} and~\eqref{2.11}
which was discarded in~\cite{Khe82} as nonphysical (because of
singularities in 2BDE) reproduce well in our case general
properties of light meson spectra. In particular, the slope
$\sigma=ka$ of light meson trajectories f\/it well to the
experimental value  if the parameter $a$ is taken from the
nonrelativistic potential model \cite{LSG91}. The third linear
potential of~\cite{Khe82} (with no singularities in 2BDE) does not
match to experimental data.

\subsection*{Acknowledgements}

The author would like to thank Professors V.~Tretyak and
I.~Simenog, Dr.~Yu.~Yaremko and Referees for helpful suggestions
and critical comments.

\LastPageEnding


\newpage


\begin{thebibliography}{99}

\footnotesize

\bibitem{Bre29}
Breit G., The ef\/fect of retardation on the interaction of two
electrons, {\it Phys. Rev.}, 1929, V.34,   553--573.

\bibitem{B-K85}
Barut A.O., Komy S., Derivation of nonperturbative reativistic
two-body equation from the action principle in quantum
electrodynamics, {\it Fortschr. Phys.}, 1985, V.33,  309--318.

\bibitem{B-U87}
Barut A.O., \"Unal N., A new approach to bound-state quantum
electrodynamics, {\it Phys. A}, 1987, V.142,  467--487.

\bibitem{Gra91}
Grandy W.T.Jr., Relativistic quantum mechanics of leptons and
f\/ields, Dordrecht~-- Boston~-- London, Kluwer Academic
Publishers, 1991.

\bibitem{D-L96}
Darewych J.W.. Di Leo L., Two-fermion Dirac-like eigenstates of
the Coulomb QED Hamiltonian, {\it J.~Phys.~A: Math. Gen.}, 1996,
V.29,
  6817--6841.

\bibitem{Dar98}
Darewych J.W., Few-particle eigenstates in the Yukawa model, {\it
Condensed Matter Physics}, 1998, V.1, N~3(15),   593--604.

\bibitem{D-D02}
Darewych J.W., Duviryak A., Exact few-particle eigenstates in
partially reduced QED, {\it Phys. Rev.~A}, 2002, V.66, 032102, 20
pages; nucl-th/0204006.

\bibitem{D-D05}
Duviryak A., Darewych J.W., Variational wave equations of two
fermions interacting via scalar, pseudoscalar, vector,
pseudovector and tensor f\/ields, {\it Cent. Eur. J. Phys.}, 2005,
V.3, N~3,   1--17.

\bibitem{F-N90}
Fushchich W.I., Nikitin A.G., On the new constants of motion for
two- and three-particle equations, {\it J.~Phys.~A: Math. Gen.},
1990, V.23, L533--L535.

\bibitem{N-F91}
Nikitin A.G., Fushchich W.I., Non-Lie integrals of the motion for
particles of arbitrary spin and for systems of interacting
particles, {\it Teor.~Mat.~Fiz.}, 1991, V.88,  406--515 (English
transl.: {\it  Theor. Math. Phys.}, 1991, V.88, 960--967).

\bibitem{S-T01}
Simenog I.V., Turovsky A.I., A relativistic model of the
two-nucleon problem with direct interaction, {\it Ukra\"\i{}n.
Fiz. Zh.}, 2001, V.46,  391--401 (in Ukrainian).

\bibitem{S-T04}
Simenog I.V., Turovsky A.I., The model of deuteron in Dirac--Breit
approach with direct interaction, {\it J.~Phys. Studies}, 2004,
V.8,  23--34 (in Ukrainian).

\bibitem{Kro76}
Krolikowski W., Relativistic radial equations for 2 spin-1/2
particles with a static interaction, {\it Acta Phys. Polon.~B},
1976, V.7, 485--496.

\bibitem{Chi87}
Childers R.W., Ef\/fective Hamiltonians for generalized Breit
interactions in QCD, {\it Phys. Rev.~D}, 1987, V.36, 606--614.

\bibitem{Bra87}
Brayshaw D.D., Relativistic description of quarkonium, {\it Phys.
Rev.~D}, 1987, V.36, 1465--1478.

\bibitem{Tsi00}
Tsibidis G.D., Quark-antiquark bound states and the Breit
equation, {\it Acta Phys. Polon. B}, 2004, V.35, 2329--2366;
hep-ph/0007143.

\bibitem{Khe82}
Khelashvili A.A., Radial quasipotential equation for a fermion and
antifermion and inf\/initely rising central potentials, {\it
Teor.~Mat.~Fiz.},
 1982, V.51,   201--210 (English transl.: {\it  Theor. Math. Phys.}, 1982, V.51,
447--453).

\bibitem{CWW96}
Crater H.W., Wong C.W. and Wong C.-Y., Singularity-free Breit
equation from constraint two-body Dirac equations, {\it Internat.
J. Modern Phys.~E}, 1996, V.5, 589--615; hep-ph/9603402.

\bibitem{M-S84}
Mlodinov L.D., Shatz M.P., Solving the Schr\"odinger equation with
use of $1/N$ perturbation theory, {\it J. Math. Phys.}, 1984,
V.25,  943--950.

\bibitem{IPS84}
Imbo T., Pagnamenta A. And Sukhatme U., Energy eigenstates of
spherically symmetric potentials using the shifted $1/N$
expansion, {\it Phys. Rev.~D}, 1984, V.29,  1669--1681.

\bibitem{Vak02}
Vakarchuk I.O., The $1/N$-expansion in quantum mechanics.
High-order approximations, {\it J. Phys. Studies} 2002, V.6,
46--54.

\bibitem{M-B97}
Mustafa O., Barakat T., Nonrelativistic shifted-l expansion
technique for three- and two-dimensional Schr\"odinger equation,
{\it Commun. Theor. Phys.}, 1997, V.28,
 257--264; math-ph/9910040.

\bibitem{M-B98}
Mustafa O., Barakat T., Relativistic shifted-l expansion technique
for Dirac and Klein--Gordon equations, {\it Commun. Theor. Phys.},
1998, V.29, 587--594; math-ph/9910039.

\bibitem{Tod71}
Todorov I.T., Quasipotential equation correspondong to the
relativistic eiconal approximation, {\it Phys. Rev.~D}, 1971, V.3,
2351--2356.

\bibitem{RST85}
Rizov V.A., Sazdian H., Todorov I.T., On the relativistic quantum
mechanics of two interacting spinless particles, {\it Ann. of
Phys. (NY)}, 1985, V.165,  59--97.

\bibitem{Duv02}
Duviryak A., Heuristic models of two-fermion relativistic systems
with f\/ield-type interaction, {\it J. Phys. G}, 2002, V.28,
2795--2809; nucl-th/0206048.

\bibitem{LSG91}
Lucha W., Schoberl F.F., Gromes D., Bound states of quarks, {\it
Phys. Rep.}, 1991, V.200, Issue 4, 127--240.

\bibitem{Edd29}
Eddington A.S.,  The charge of an electron, {\it  R. Soc. Lond.
Proc. Ser. A}, 1929, V.122, N~789, 358--369.

\bibitem{Gau29}
Gaunt J.A., The triplets of Helium,
{\it  Philos. Trans. R. Soc. Lond. Ser. A}, 1929, V.228, 151--196.\\
Gaunt J.A., The triplets of Helium,  {\it R. Soc. Lond. Proc. Ser. A}, 1929, V.122,
N~790,   513--532.

\bibitem{Sal52}
Salpeter E.E., Mass corrections to the f\/ine structure of
hydrogen-like atoms, {\it Phys. Rev.}, 1952, V.87,   328--343.

\bibitem{Fau66}
Faustov R.N., The proton structure and hyperf\/ine splitting of
hydrogen energy levels, {\it Nucl. Phys.}, 1966, V.75, 669--681.

\bibitem{Khe69}
Khelashvili A.A., Quasipotential equation for the system of two
particles with spin 1/2, {\it Communications of the Joint
Institute for Nuclear Physics}, P2--4327, Dubna, 1969 (in
Russian).

\bibitem{L-R83}
Long C., Robson D., Bound states of a relativistic quark
conf\/ined by a vector potential, {\it Phys. Rev. D}, 1983, V.27,
644--646.

\bibitem{B-J82}
Baric N., Jena S.N., Lorentz structure vs relativistic consistency
of an ef\/fective power-law potential model for quark-antiquark
systems, {\it Phys. Rev.~D}, 1982, V.26, 2420--2429.

\bibitem{HLS96}
Haysak I., Lengyel V., Shpenik A., Challupka S., Salak M., Quark masses in the relativistic
analytic model, {\it Ukra\"\i{}n. Fiz. Zh.}, 1996, V.41, 370--372
(in Ukrainian).
\end{thebibliography}
\end{document}